\documentclass[%
 reprint,
superscriptaddress,
 aps,
 prl
]{revtex4-2}

\usepackage{xfrac}
\usepackage{braket}
\usepackage{physics}
\usepackage{graphicx}
\usepackage{dcolumn}
\usepackage{bm}
\usepackage{subcaption}
\DeclareMathSizes{10}{10}{4}{4}
\usepackage[colorlinks=true, allcolors=blue]{hyperref}
\usepackage{floatrow}
\newcommand{\labelphantom}[1]{%
  \parbox{0pt}{\phantomsubcaption\label{#1}}%
}
\newcommand{\eqn}[1]{Eq.~(\ref{#1})}
\newcommand{\Eqn}[1]{Equation~(\ref{#1})}

\usepackage{natbib}
\usepackage[resetlabels]{multibib}
\newcites{SI}{SI references}

\begin{document}
\preprint{APS/123-QED}

\title{Instantons and the quantum
bound to chaos}

\author{Vijay Ganesh Sadhasivam}\email{vgs23@cam.ac.uk}
\affiliation{Yusuf Hamied Department of Chemistry, University of Cambridge, Lensfield Road, Cambridge, CB2 1EW, UK}

\author{Lars Meuser}
\affiliation{Yusuf Hamied Department of Chemistry, University of Cambridge, Lensfield Road, Cambridge, CB2 1EW, UK}
\affiliation{Laboratory of Physical Chemistry, ETH Zürich, 8093 Zürich, Switzerland}
\author{David R.\ Reichman}
\affiliation{Department of Chemistry, Columbia University, New York, New York 10027, USA}

\author{Stuart C.\ Althorpe}\email{sca10@cam.ac.uk}
\affiliation{Yusuf Hamied Department of Chemistry, University of Cambridge, Lensfield Road, Cambridge, CB2 1EW, UK}

\date{\today}

\begin{abstract}
We investigate why the Lyapunov exponents $\lambda$ of out-of-time-ordered correlators (OTOCs) satisfy a universal bound $\lambda < 2 \pi k_B T/{\hbar}$ by probing imaginary-time path-integral space using ring-polymer molecular dynamics (RPMD) for a barrier-crossing model. We find that the RPMD OTOC satisfies the same bound as the quantum OTOC, which is caused by the stability of quantum thermal fluctuations around the instanton on the barrier. We expect that similar instantons (or other delocalised structures) will be found in many other systems with exponentially growing OTOCs.
\end{abstract}

\maketitle

`Quantum scrambling' refers to the tendency of an initially localised quantum state to lose information by spreading across the system \cite{swingle2018unscrambling}. There has been much interest recently in quantifying the rate of quantum scrambling using out-of-time-ordered correlators (OTOCs) \cite{maldacena2016bound,sekino2008fast,larkin1969quasiclassical,rozenbaum2017lyapunov,xu2020does} which have the general form
\begin{equation}
    C(t) = \langle  [\hat{W}(t), \hat{V}(0) ]^{\dagger}  [\hat{W}(t), \hat{V}(0) ] \rangle
\end{equation}
where $\hat V$ and $\hat W$ are Hermitian operators, and $\langle\rangle$ denotes a thermal average. Of particular interest are systems which are said (loosely) to exhibit `quantum chaos', which in this context means that the OTOC grows (after a short transient time) as
\begin{equation}
    C(t) \sim a\epsilon \exp({{\lambda}t})
\end{equation}
before the growth stops at some `Ehrenfest time', by which the system has explored sufficient Hilbert space that real-time coherence effects dominate the OTOC;  $\lambda$ is often referred to as the `quantum Lyapunov exponent', since it tends to the classical Lyapunov exponent in the classical limit (where the small parameter $\epsilon\sim\hbar^2$). 

In 2016, Maldacena et al.\ proved \cite{maldacena2016bound} (under certain assumptions) that $\lambda$ satisfies the bound
\begin{align}\label{MSSbound}
    \lambda \leq \frac{2\pi k_B T}{\hbar}.
\end{align}
This remarkable result has attracted a lot of interest \cite{swingle2018unscrambling,rozenbaum2017lyapunov,xu2020does,blake2021systems,UniversalOpgrowth,RegOTOC}, since it implies a universal limit to the rate at which a quantum system can scramble information. \Eqn{MSSbound} has been verified for a wide variety of model systems, ranging from few-dimensional models  \cite{DickeModelOTOC,pappalardi2020quantum,MoleculeOTOC}, to many-body fermionic systems, some of which, such as the Sachdev-Ye-Kitaev model \cite{sachdev1993gapless,KitaevSYK}, have been shown to saturate the bound. To our knowledge, no unifying mechanism has been proposed to explain how the bound is imposed in such a diverse range of systems, nor why it takes such a simple, system-independent form. However, a growing body of work \cite{murthy2019bounds, tsuji,Pappalardi2022,nussinov2022exact} suggests that the bound is of statistical origin. In particular, Tsuji et al.\ \cite{tsuji} have re-derived the bound from the Kubo-Martin Schwinger (KMS) relation (which allowed a factorization assumption made in the original derivation to be relaxed), and Pappalardi et al.\ \cite{Pappalardi2022} have shown that the bound of \eqn{MSSbound} is related to a similar bound noted previously in quantum transport.

The KMS relation is a very strict condition which is satisfied (to our knowledge) only by the exact quantum dynamics. This raises the question whether the statistical conditions necessary to impose the bound might be less restrictive. In particular, it is easy, by using the discretised Euclidean action as a potential energy surface, to generate an artificial classical dynamics in an extended phase space which conserves the quantum Boltzmann distribution \cite{rahman,voth,craig2004quantum}. Would this condition alone be sufficient to make $\lambda$ satisfy the bound (when the artificial classical dynamics is chaotic), and if so, does the (artificial) classical dynamics yield insight into how the statistics imposes the bound?

\begin{figure}
    \centering
    \labelphantom{fig1a}%
    \labelphantom{fig1b}%
    \labelphantom{fig1c}%
    \labelphantom{fig1d}
    \includegraphics[width=0.95\linewidth]{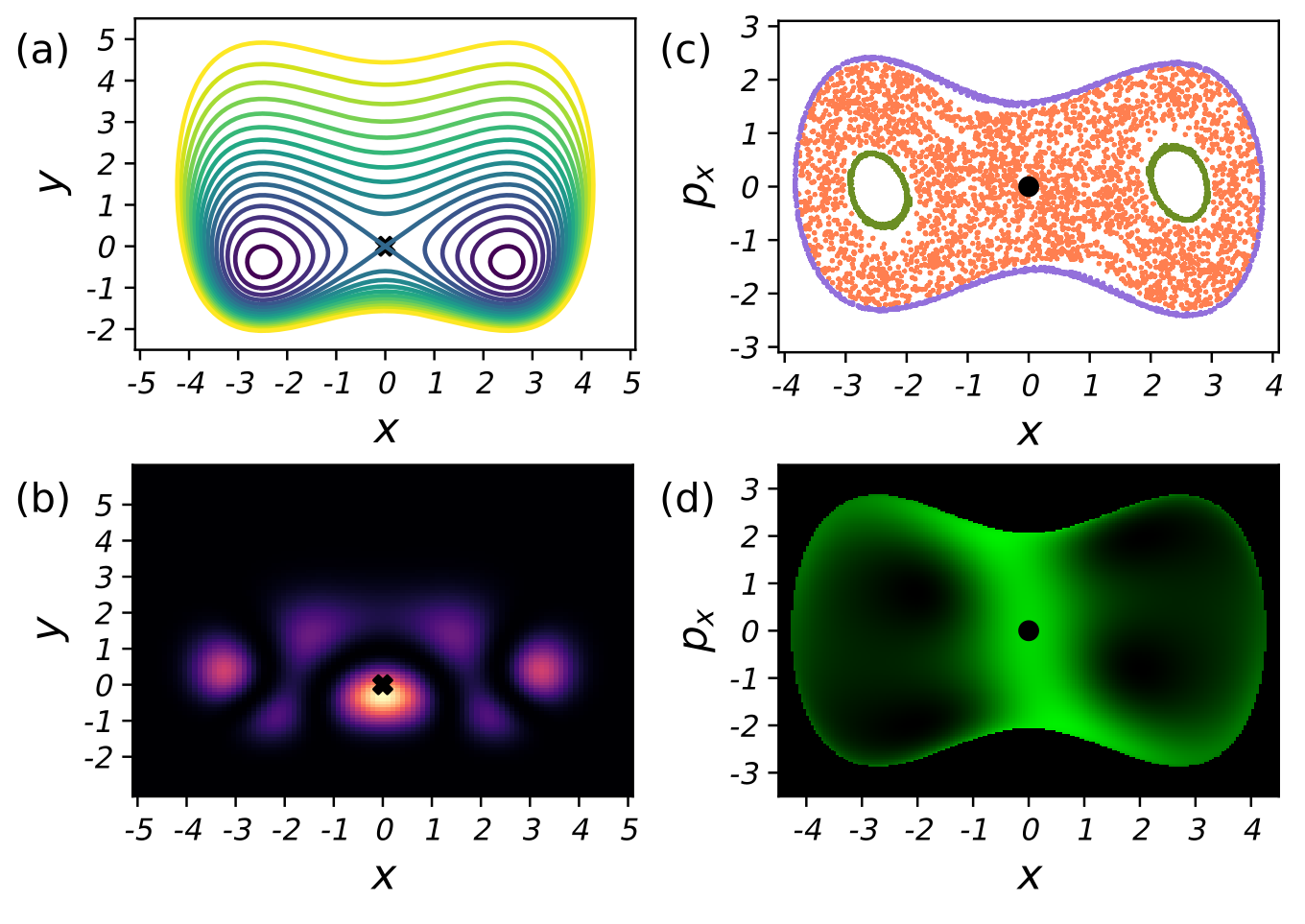}
    \caption{Characterisation of the double-well model used to calculate the results in Figs.~2 and 3, showing (a) the potential energy surface, (b) state-mixing in the eigenstate $\ket{n_x = 3, n_y = 2}$ for which the energy is just above the barrier top, (c) and (d) the $\text{Poincar\'e}$ and Husimi sections at the same energy (with $y=0$).}
    \label{fig1}
\end{figure}

To address these questions, we report here ring-polymer molecular dynamics (RPMD) calculations for a classically chaotic double-well model. We use RPMD because it is the cheapest available method for propagating classical trajectories that conserves the quantum Boltzmann distribution \cite{craig2004quantum, annurev}. The double-well hamiltonian is
\begin{align}\label{wells}
H=& {{\bf p}^2\over 2m}+g\left(x^2 - \frac{m\omega_b^2}{4g}\right)^2 + D(1-e^{-\alpha y})^2 \nonumber\\&- gx^2\left (x^2-\frac{m\omega_b^2}{2g} \right)(1-e^{-\alpha y})
\end{align}
(with $m=0.5$, $g=0.08$, $\omega_b = 2$, $D=3V_b$, $V_b=m^2\omega_b^4/16g$ and $\alpha=0.382$). These parameters were chosen to give a system in which barrier crossing is not a rare event, even at temperatures sufficiently low for strong quantum effects (see below). Classically, the barrier crossing gives rise to separatrix chaos (see Fig.~1c), and quantum mechanically, to barrier scrambling, whereby a state initially on one side of the barrier delocalises over both sides. Previous studies \cite{hashimoto2020exponential} have shown that such scrambling gives rise to an OTOC which grows exponentially.

The RPMD trajectories were propagated in the $4N$-dimensional `ring-polymer' phase space obtained by replicating the system $N$ times, using the hamiltonian
\begin{align}\label{RPHamil}
    H_N(\textbf{q},\textbf{p}) = \sum_{i=1}^{N} \frac{\textbf{p}_i\cdot\textbf{p}_i}{2m} + U_N(\textbf{q})
\end{align} in which $m$ is the physical mass, and
\begin{align}\label{RP-PE}
    U_N(\textbf{q}) = \sum_{i=1}^N V(\textbf{q}_i) +  \frac{m}{2(\beta_N\hbar)^2} \sum_{i=1}^N  \vert \textbf{q}_i - \textbf{q}_{i-1} \vert^2
\end{align}
is the Euclidean action, discretised into $N$ imaginary-time steps $\beta_N\equiv\beta/N$, with $\beta=1/k_\text{B}T$. It is easy to show that 
\begin{align}
    \left\langle A(\hat {\bf q})\right\rangle =\lim_{N\to\infty} \frac{1}{\mathcal{Z}}\int d\textbf{p}\! \int d\textbf{q}\; e^{-\beta H_N(\textbf{p},\textbf{x})} A({\bf q})
\end{align}
(where $\mathcal{Z}/h^{2N}$ is the partition function) and in this respect RPMD can be regarded as a variety of path-integral molecular dynamics \cite{rahman}, in which the RPMD trajectories are simply a tool for sampling imaginary-time path-integral space. However, the choice of $m$ as the physical mass ensures that RPMD gives thermal time-correlation functions (TCFs) which agree with the exact quantum Kubo TCFs up to some power of the time $t$ \cite{braams}, and this property can  be used also to  define (for $\hat W=\hat x$,  $\hat V=\hat p_x$) an RPMD OTOC,
\begin{align}\label{rpmdo}
    C^\text{RP}_N(t) = \frac{\hbar^2}{\mathcal{Z}}\int d\textbf{p}^N\! \int d\textbf{q}^N\; e^{-\beta_N H_N(\textbf{p},\textbf{q})} \left \vert \frac{\partial X_{0t}}{\partial X_0} \right \vert^2
\end{align}
in which $\partial X_t/{\partial X_0}$ is the stability matrix of the ring-polymer centroid coordinate 
\begin{align}
X={1\over N}\sum_{i=1}^Nx_i.
\end{align}
It is straightforward to show that $C^\text{RP}_N(t)$ and the quantum Kubo OTOC
\begin{align}\label{kubo}
C(t)={1\over\beta\mathcal{Z}}\int^\beta_0\!d\gamma\,\text{Tr}\left\{e^{-\gamma\hat H}[\hat x(t),\hat p_x]^\dag e^{-(\beta-\gamma)\hat H} [\hat x(t),\hat p_x]\right\}
\end{align}
are both even functions of $t$, and that they agree at short times to within a leading order error of ${\cal O}(t^6)$. One can also show (by collapsing the beads) that $C^\text{RP}_N(t)$ tends to the classical limit of the OTOC,
 \begin{align}
    C_T^{cl}(t) = \frac{\hbar^2}{\mathcal{Z}_{cl}}\int d\textbf{p}\! \int d\textbf{q}\; e^{-\beta H_{cl}(\textbf{p},\textbf{q})} \left \vert \frac{\partial x_t}{\partial x_0} \right \vert^2
\end{align}
as $T\to\infty$. We thus expect the RPMD OTOC to agree with the quantum Kubo OTOC for very short times, and
with the classical OTOC at all times at high temperatures, but otherwise it should be regarded as a classical OTOC in an artificially extended phase space, with the crucial property that its dynamics conserves the quantum Boltzmann distribution (of the real system).

The resulting OTOCs, calculated for the system of \eqn{wells}, are shown in Fig.~2a. The RPMD and classical stability matrix elements $dx_t/dx_0$ and $dX_t/dX_0$ were propagated using the fourth-order symplectic propagator of ref.~\cite{brewer1997semiclassical}; all other numerical details were standard (see the Appendix).  As expected, the RPMD OTOCs match the quantum Kubo OTOCs at short times, and line up with the classical OTOCs at high temperatures ($>2 T_c$, with $T_c$ defined below); but otherwise they disagree with the quantum OTOCs, showing a later onset of exponential growth and (also as expected) no flattening of the curve after the Ehrenfest time \footnote{The oscillations in the post-Ehrenfest part of the quantum OTOC at $0.95T_c$ demonstrate that the system has not completely thermalised after scrambling. Similar differences in the scrambling and thermalization timescales have been found in calculations on the Dicke model \cite{DickeModelOTOC}.}. However, the RPMD OTOCs do show exponential growth over the same ($0.7T_c$ to $3T_c$) temperature range as the quantum OTOCs, because the quantum Boltzmann statistics allows the RPMD trajectories access to the barrier-top (which, classically, becomes a rare-event below about $1.8T_c$).

\begin{figure*}
    \includegraphics[width=\linewidth]{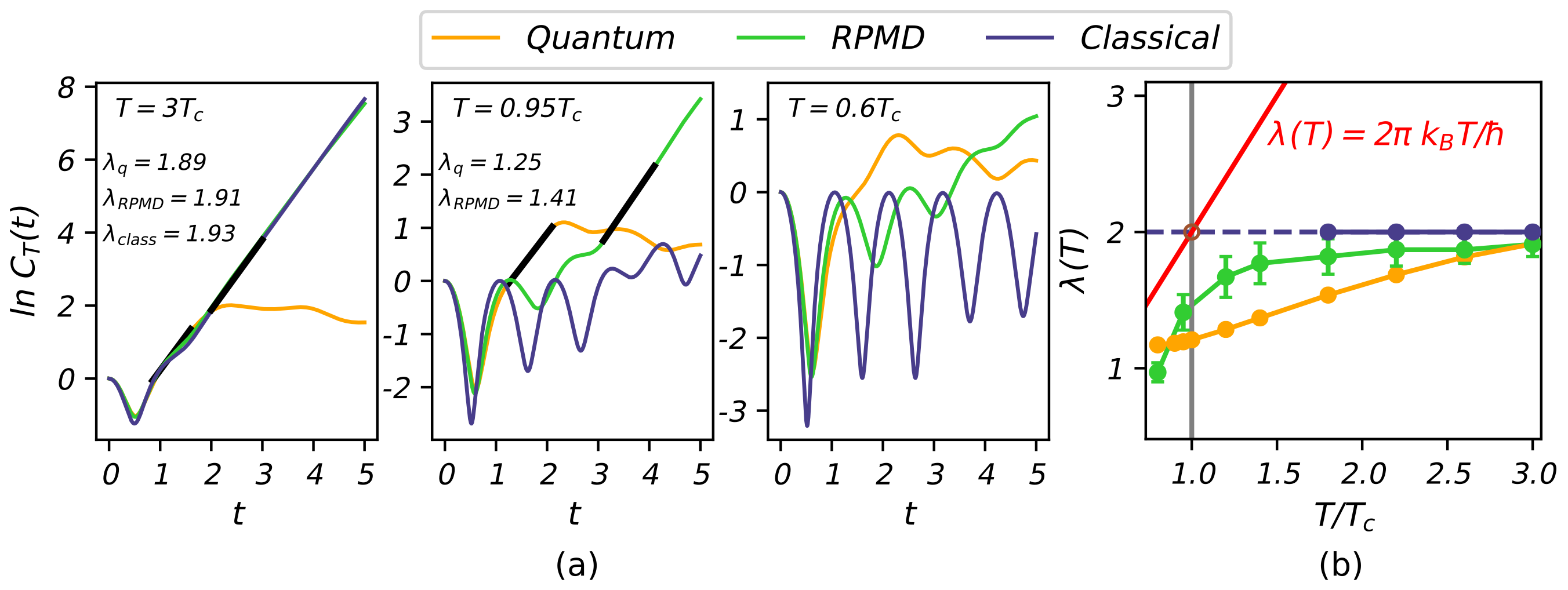}
    \caption{(a) Out-of-time ordered correlators (OTOCs) computed for the double-well system of Fig.~1. (b) Temperature dependence of the Lyapunov exponent $\lambda$ extracted from the corresponding OTOCs, with the bound of \eqn{MSSbound} shown in red. }
    \label{fig3}
\end{figure*}

Figure 2b plots the temperature dependence of the values of $\lambda$ obtained from the exponential growth stages of the various OTOCs. As expected, the quantum $\lambda$ is close to the classical $\lambda$  (which is simply the barrier-crossing frequency $\omega_b$) at $T>2T_c$, but then decreases as the temperature is reduced in order to satisfy the bound (red line) \footnote{Note that the quantum Kubo OTOC of \eqn{kubo} is different from the various types of symmetrised quantum OTOC usually considered when discussing the bound \cite{RegOTOC}. However, like the symmetrised OTOCs, the quantum Kubo OTOC is `regularised', with the two commutators separated by ${\exp}(-\gamma \hat H)$.}.  Not surprisingly (given the differences in the OTOC plots of Figs.~2a)   the RPMD $\lambda$ is different from the quantum $\lambda$, and for most temperatures below 1.8$T_c$ is significantly higher. However, the key result of Fig.~2b is that  the {\em RPMD OTOC satisfies the bound of \eqn{MSSbound}}. Since the exponential growth occurs at times significantly later than the short timespan over which the quantum and RPMD OTOCs agree, the fact that RPMD obeys the bound is either a coincidence, or a consequence of the quantum Boltzmann statistics (which the RPMD trajectories conserve).

A clue that it is the statistics that impose the bound is that the RPMD $\lambda$ bends downwards at temperatures near the `cross-over temperature'
\begin{align}\label{cross}
T_c = {\hbar\omega_b\over2\pi k_\text{B}}
\end{align}
below which the Boltzmann distribution at the barrier top becomes dominated by thermal fluctuations around an instanton \cite{miller,benders,richardson2009ring}.  The instanton forms when  the frequency of the first Matsubara mode $\sqrt{4\pi^2/(\beta\hbar)^2-\omega_b^2}$ becomes imaginary, causing the saddle point on $U({\bf x})$ to move from the classical geometry at which all the ring-polymer `beads' lie on the saddle point of $V(x,y)$,  to the instanton geometry ${\bf \tilde x}$, in which the beads occupy $N$ discrete time-intervals along an unstable periodic orbit (with period $\beta\hbar$) on the inverted potential $-V(x)$.  The formation of the instanton at $T_c$ implies that classical barrier-scrambling cannot take place at $T<T_c$, and Hashimoto and Tanahashi \cite{HawkingTempHashimoto} have pointed out  that such a restriction is consistent with the bound, since the classical value of $\lambda$ cannot exceed $\omega_b$. 
 
The exponential growth of the RPMD OTOC, however, extends to temperatures significantly below $T_c$.  
Let us make the assumption (which we test below) that the exponential growth of the RPMD OTOC at $T<T_c$ is caused by trajectories which are chaotic because they pass close to the instanton. We would then expect the RPMD $\lambda$ to be limited by $\eta$, analogously to how the classical $\lambda$ is limited by $\omega_b$, where
\begin{align}
\eta^2 = -{1\over m}{\partial^2 U({\bf \tilde x})\over \partial X_0^2}=-{1\over mN}\sum_i{\partial^2 V(\tilde x_i,\tilde y_i)\over \partial x_i^2}
\end{align}
which is the projection of the ring-polymer Hessian along the centroid coordinate $X_0$ (since the RPMD OTOC of \eqn{rpmdo} follows the exponential growth of $\abs{\partial X_{0t} / \partial X_0}^2$).
The Hessian eigenvalues at the instanton geometry are known to be all positive, except for a single unstable mode, and a zero-frequency mode, corresponding to imaginary-time translation \cite{richardson2009ring}. Since ${\partial^2 U({\bf \tilde x})/ \partial X_0^2}<0$, it follows that the Hessian along any mode orthogonal to the centroid is positive, and hence that
\begin{align}\label{bo11}
{\partial^2 U({\bf \tilde x})\over \partial X_1^2}+ {\partial^2 U({\bf \tilde x})\over \partial X_{\overline 1}^2}\ge 0
\end{align}
where $\{X_1,X_{\overline 1}\}$ are the two lowest frequency Fourier modes ($\sqrt{2/ N}\sum_{i=1}^N\{\sin,\cos\}\left(2\pi i/N \right)x_i$) orthogonal to the centroid.
 Substituting for $U({\bf \tilde x})$ in \eqn{bo11} using \eqn{RP-PE} and noting that each of $X_1$ and $X_{\overline 1}$ diagonalizes $S({\bf x})$ to give $2m\pi^2/(\beta\hbar)^2$ (in the limit $N\to\infty$), we obtain,
\begin{align}
\eta\le&\frac{2\pi k_B T}{\hbar}
\end{align}
which implies that the RPMD OTOC must satisfy the bound if all the chaotic trajectories pass close to the instanton geometry.

\begin{figure*}[!htb]
    \centering
    \labelphantom{fig4a}
    \labelphantom{fig4b}
    \labelphantom{fig4c}
    \labelphantom{fig4d}
    \labelphantom{fig4e}
        \includegraphics[width=\linewidth]{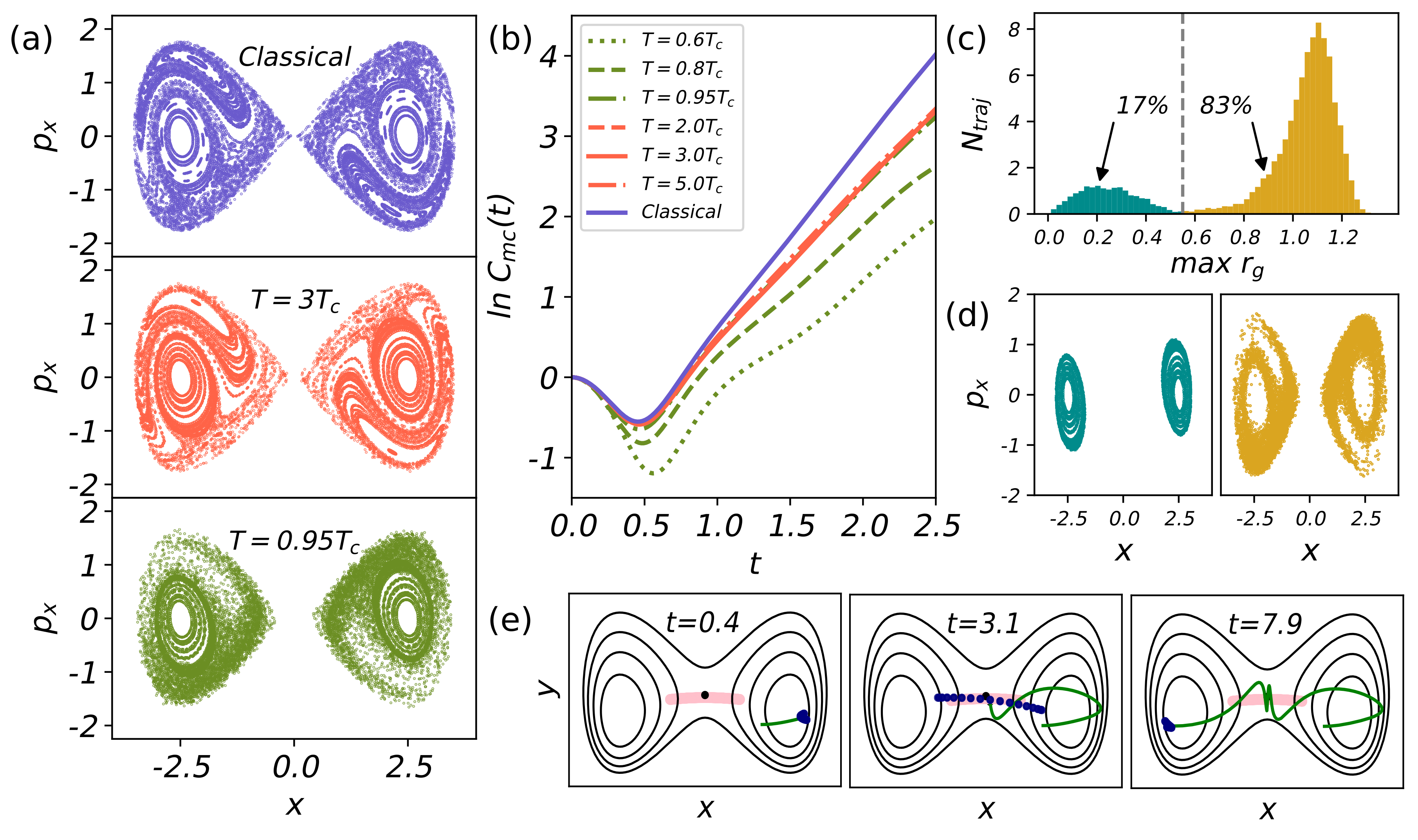}
    \caption{Analysis of the RPMD trajectories, showing  (a) Poincar\'e surfaces of section obtained classically and for the RPMD centroids (as described in the text); (b) `microcanonical' RPMD OTOCs at various temperatures obtained from \eqn{micro} and compared with the classical microcanonical OTOC; (c) histogram of the maximum radius of gyration along the ring-polymer trajectories contributing to the 0.95$T_c$ RPMD Poincar\'e section; (d) decomposition of the same section into compact and instantonic ring-polymer trajectories; (e)  example of a chaotic RPMD trajectory passing close to the instanton.}
    \label{fig:my_label}
\end{figure*}

To verify that the chaotic RPMD trajectories do pass close to the instanton, we decompose the thermal RPMD OTOCs into the `microcanonical' RPMD OTOCs \footnote{These OTOCs are of course not microcanonical except in the high-temperature limit (since $H_N$ is not the energy); they are used here simply to probe the dynamics of the trajectories with energy close to the barrier top.}
\begin{align}\label{micro}
    C^\text{micro}_N(t) = \frac{\hbar^2}{\mathcal{Z}(E)}\int d\textbf{p}^N\! \int d\textbf{q}^N\; \delta[H_N({\bf p},{\bf q})\!-\!NV_b] \left \vert \frac{\partial X_t}{\partial X_0} \right \vert^2
\end{align}
at several temperatures (see Fig.~3b). These OTOCs also grow exponentially, with values of $\lambda$ which decrease rapidly as $T$ drops below $T_c$ and which are greater than the corresponding thermal values. We analyse the trajectories using `centroid  Poincar\'e surfaces of section', obtained by fixing  $E=NV_b$ and $Y_t=0$, and plotting $(X_t,P_{xt})$ while integrating over the other non-centroid degrees of freedom in \eqn{micro}. The centroid section is very close to the classical Poincar\'e section (with $E_{\rm class}=V_b$, $y=0$) at $3T_c$ (see Fig.~3a), but rather different at $0.95T_c$, where the polymers are floppier, with far fewer phase-space points that can be identified as lying along integrable trajectories. A histogram of the maximum radius of gyration of the ring polymer trajectories contributing to the $0.95T_c$ centroid Poincar\'e section is bimodal (Fig.~3c), indicating that some polymer geometries correspond to thermal quantum fluctuations around classical geometries (in which all the polymer beads are in the same place), others to fluctuations around the instanton (the pink line in Fig.~3e). Figure 3d shows that the former contribute only to the integrable trajectories in the centroid Poincar\'e section. The scrambling trajectories, responsible for the exponential growth of the OTOC, must therefore all pass through thermal fluctuations around the instanton; an example of such a trajectory is given in Fig.~3e (see also the supplementary material for examples of longer trajectories). 

Thus the RPMD $\lambda$ of Fig.~2b obeys the bound as a direct consequence of the simple properties of the imaginary-time action that give \eqn{bo11}. Expressed in general terms, these properties are that, if exponential growth in the RPMD OTOC is to survive cooling, then the unstable chaos-producing features of the classical potential must transform (on cooling) to analogous features of the imaginary-time action involving the centroid degrees of freedom; the Matsubara fluctuations around the centroids will be stable, and it is this stability that imposes the bound on the RPMD $\lambda$, through \eqn{bo11}. These properties are sufficiently general that one can imagine them imposing the bound in an analogous way for systems in which the chaos is produced by mechanisms very different from barrier scrambling; but clearly this hypothesis will need to be tested numerically.

Another question for future studies is whether this simple idea of the origin of the bound can be applied directly to the exact quantum dynamics. For the double-well system considered here we can make analogies with chemical reactions, for which the flux of the RPMD trajectories close to the instanton
 is known to give a good estimate of the exact quantum flux over the barrier \cite{richardson2009ring,tim}. However, unlike a chemical reaction, a barrier-scrambling system (with an exponentially growing OTOC) has a low barrier, and this permits real-time coherent tunnelling to make a large contribution to the exponential growth of the quantum OTOC (as is easily verified by decomposing the OTOC into contributions from individual eigenstates). Coherent tunnelling is obviously missing from RPMD, which we think explains both why the onset of exponential growth is delayed in the RPMD OTOCs of Fig.~2a (since it takes longer for the barrier-crossing dynamics to grow sufficiently to dominate the OTOC) and why the RPMD $\lambda$ is greater than the exact quantum $\lambda$ (because the coherent tunnelling contributes smaller values of $\lambda$ than the incoherent tunnelling at the barrier top). This suggests  that the OTOC bound applies only to incoherent scrambling dynamics, and thus may perhaps fail in some systems for which real-time coherence dominates the OTOC at short times.
 
This article has focused on a simple model with quantum Boltzmann statistics. We think it likely that analogous instantons will also be found in the wide variety of bosonic and fermionic models for which exponentially growing OTOCs have been identified \cite{SPscramNotherm,kidd2020thermalization,fu2016numerical}, although such instantons may be difficult to find (since they would require formulation of the OTOC as a coherent-state path-integral and would live in the complex plane). Thus the most promising future use of the simple tools introduced above is likely to be in exploring a wide range of quantum Boltzmann systems, using  RPMD  as a black-box method to find exponentially growing OTOCs and identify the associated instantons. An interesting target for such calculations would be to search for examples of quantum Boltzmann systems which saturate the bound.

We would like to thank Srihari Keshavamurthy, Yair Litman, George Trenins, Jeremy Richardson and Venkat Kapil for stimulating discussions and for commenting on an earlier version of this manuscript. VGS acknowledges the support of a Manmohan Singh PhD scholarship from St John's College, Cambridge.





\bibliography{Citations}

\begin{thebibliography}{38}%
\makeatletter
\providecommand \@ifxundefined [1]{%
 \@ifx{#1\undefined}
}%
\providecommand \@ifnum [1]{%
 \ifnum #1\expandafter \@firstoftwo
 \else \expandafter \@secondoftwo
 \fi
}%
\providecommand \@ifx [1]{%
 \ifx #1\expandafter \@firstoftwo
 \else \expandafter \@secondoftwo
 \fi
}%
\providecommand \natexlab [1]{#1}%
\providecommand \enquote  [1]{``#1''}%
\providecommand \bibnamefont  [1]{#1}%
\providecommand \bibfnamefont [1]{#1}%
\providecommand \citenamefont [1]{#1}%
\providecommand \href@noop [0]{\@secondoftwo}%
\providecommand \href [0]{\begingroup \@sanitize@url \@href}%
\providecommand \@href[1]{\@@startlink{#1}\@@href}%
\providecommand \@@href[1]{\endgroup#1\@@endlink}%
\providecommand \@sanitize@url [0]{\catcode `\\12\catcode `\$12\catcode
  `\&12\catcode `\#12\catcode `\^12\catcode `\_12\catcode `\%12\relax}%
\providecommand \@@startlink[1]{}%
\providecommand \@@endlink[0]{}%
\providecommand \url  [0]{\begingroup\@sanitize@url \@url }%
\providecommand \@url [1]{\endgroup\@href {#1}{\urlprefix }}%
\providecommand \urlprefix  [0]{URL }%
\providecommand \Eprint [0]{\href }%
\providecommand \doibase [0]{https://doi.org/}%
\providecommand \selectlanguage [0]{\@gobble}%
\providecommand \bibinfo  [0]{\@secondoftwo}%
\providecommand \bibfield  [0]{\@secondoftwo}%
\providecommand \translation [1]{[#1]}%
\providecommand \BibitemOpen [0]{}%
\providecommand \bibitemStop [0]{}%
\providecommand \bibitemNoStop [0]{.\EOS\space}%
\providecommand \EOS [0]{\spacefactor3000\relax}%
\providecommand \BibitemShut  [1]{\csname bibitem#1\endcsname}%
\let\auto@bib@innerbib\@empty
\bibitem [{\citenamefont {Swingle}(2018)}]{swingle2018unscrambling}%
  \BibitemOpen
  \bibfield  {author} {\bibinfo {author} {\bibfnamefont {B.}~\bibnamefont
  {Swingle}},\ }\bibfield  {title} {\bibinfo {title} {Unscrambling the physics
  of out-of-time-order correlators},\ }\href@noop {} {\bibfield  {journal}
  {\bibinfo  {journal} {Nat. Phys.}\ }\textbf {\bibinfo {volume} {14}},\
  \bibinfo {pages} {988} (\bibinfo {year} {2018})}\BibitemShut {NoStop}%
\bibitem [{\citenamefont {Maldacena}\ \emph {et~al.}(2016)\citenamefont
  {Maldacena}, \citenamefont {Shenker},\ and\ \citenamefont
  {Stanford}}]{maldacena2016bound}%
  \BibitemOpen
  \bibfield  {author} {\bibinfo {author} {\bibfnamefont {J.}~\bibnamefont
  {Maldacena}}, \bibinfo {author} {\bibfnamefont {S.~H.}\ \bibnamefont
  {Shenker}},\ and\ \bibinfo {author} {\bibfnamefont {D.}~\bibnamefont
  {Stanford}},\ }\bibfield  {title} {\bibinfo {title} {A bound on chaos},\
  }\href@noop {} {\bibfield  {journal} {\bibinfo  {journal} {J. High Energy
  Phys.}\ }\textbf {\bibinfo {volume} {2016}}\bibinfo  {number} { (8)},\
  \bibinfo {pages} {1}}\BibitemShut {NoStop}%
\bibitem [{\citenamefont {Sekino}\ and\ \citenamefont
  {Susskind}(2008)}]{sekino2008fast}%
  \BibitemOpen
\bibfield  {number} {  }\bibfield  {author} {\bibinfo {author} {\bibfnamefont
  {Y.}~\bibnamefont {Sekino}}\ and\ \bibinfo {author} {\bibfnamefont
  {L.}~\bibnamefont {Susskind}},\ }\bibfield  {title} {\bibinfo {title} {Fast
  scramblers},\ }\href@noop {} {\bibfield  {journal} {\bibinfo  {journal} {J.
  High Energy Phys.}\ }\textbf {\bibinfo {volume} {2008}}\bibinfo  {number} {
  (10)},\ \bibinfo {pages} {065}}\BibitemShut {NoStop}%
\bibitem [{\citenamefont {Larkin}\ and\ \citenamefont
  {Ovchinnikov}(1969)}]{larkin1969quasiclassical}%
  \BibitemOpen
\bibfield  {number} {  }\bibfield  {author} {\bibinfo {author} {\bibfnamefont
  {A.}~\bibnamefont {Larkin}}\ and\ \bibinfo {author} {\bibfnamefont {Y.~N.}\
  \bibnamefont {Ovchinnikov}},\ }\bibfield  {title} {\bibinfo {title}
  {Quasiclassical method in the theory of superconductivity},\ }\href@noop {}
  {\bibfield  {journal} {\bibinfo  {journal} {Sov. Phys. JETP}\ }\textbf
  {\bibinfo {volume} {28}},\ \bibinfo {pages} {1200} (\bibinfo {year}
  {1969})}\BibitemShut {NoStop}%
\bibitem [{\citenamefont {Rozenbaum}\ \emph {et~al.}(2017)\citenamefont
  {Rozenbaum}, \citenamefont {Ganeshan},\ and\ \citenamefont
  {Galitski}}]{rozenbaum2017lyapunov}%
  \BibitemOpen
  \bibfield  {author} {\bibinfo {author} {\bibfnamefont {E.~B.}\ \bibnamefont
  {Rozenbaum}}, \bibinfo {author} {\bibfnamefont {S.}~\bibnamefont
  {Ganeshan}},\ and\ \bibinfo {author} {\bibfnamefont {V.}~\bibnamefont
  {Galitski}},\ }\bibfield  {title} {\bibinfo {title} {Lyapunov exponent and
  out-of-time-ordered correlator’s growth rate in a chaotic system},\ }\href
  {https://doi.org/10.1103/PhysRevLett.118.086801} {\bibfield  {journal}
  {\bibinfo  {journal} {Phys. Rev. Lett.}\ }\textbf {\bibinfo {volume} {118}},\
  \bibinfo {pages} {086801} (\bibinfo {year} {2017})}\BibitemShut {NoStop}%
\bibitem [{\citenamefont {Xu}\ \emph {et~al.}(2020)\citenamefont {Xu},
  \citenamefont {Scaffidi},\ and\ \citenamefont {Cao}}]{xu2020does}%
  \BibitemOpen
  \bibfield  {author} {\bibinfo {author} {\bibfnamefont {T.}~\bibnamefont
  {Xu}}, \bibinfo {author} {\bibfnamefont {T.}~\bibnamefont {Scaffidi}},\ and\
  \bibinfo {author} {\bibfnamefont {X.}~\bibnamefont {Cao}},\ }\bibfield
  {title} {\bibinfo {title} {Does scrambling equal chaos?},\ }\href
  {https://doi.org/10.1103/PhysRevLett.124.140602} {\bibfield  {journal}
  {\bibinfo  {journal} {Phys. Rev. Lett.}\ }\textbf {\bibinfo {volume} {124}},\
  \bibinfo {pages} {140602} (\bibinfo {year} {2020})}\BibitemShut {NoStop}%
\bibitem [{\citenamefont {Blake}\ and\ \citenamefont
  {Liu}(2021)}]{blake2021systems}%
  \BibitemOpen
  \bibfield  {author} {\bibinfo {author} {\bibfnamefont {M.}~\bibnamefont
  {Blake}}\ and\ \bibinfo {author} {\bibfnamefont {H.}~\bibnamefont {Liu}},\
  }\bibfield  {title} {\bibinfo {title} {On systems of maximal quantum chaos},\
  }\href@noop {} {\bibfield  {journal} {\bibinfo  {journal} {J. High Energy
  Phys.}\ }\textbf {\bibinfo {volume} {2021}}\bibinfo  {number} { (5)},\
  \bibinfo {pages} {1}}\BibitemShut {NoStop}%
\bibitem [{\citenamefont {Parker}\ \emph {et~al.}(2019)\citenamefont {Parker},
  \citenamefont {Cao}, \citenamefont {Avdoshkin}, \citenamefont {Scaffidi},\
  and\ \citenamefont {Altman}}]{UniversalOpgrowth}%
  \BibitemOpen
\bibfield  {number} {  }\bibfield  {author} {\bibinfo {author} {\bibfnamefont
  {D.~E.}\ \bibnamefont {Parker}}, \bibinfo {author} {\bibfnamefont
  {X.}~\bibnamefont {Cao}}, \bibinfo {author} {\bibfnamefont {A.}~\bibnamefont
  {Avdoshkin}}, \bibinfo {author} {\bibfnamefont {T.}~\bibnamefont
  {Scaffidi}},\ and\ \bibinfo {author} {\bibfnamefont {E.}~\bibnamefont
  {Altman}},\ }\bibfield  {title} {\bibinfo {title} {A universal operator
  growth hypothesis},\ }\href {https://doi.org/10.1103/PhysRevX.9.041017}
  {\bibfield  {journal} {\bibinfo  {journal} {Phys. Rev. X}\ }\textbf {\bibinfo
  {volume} {9}},\ \bibinfo {pages} {041017} (\bibinfo {year}
  {2019})}\BibitemShut {NoStop}%
\bibitem [{\citenamefont {Liao}\ and\ \citenamefont
  {Galitski}(2018)}]{RegOTOC}%
  \BibitemOpen
  \bibfield  {author} {\bibinfo {author} {\bibfnamefont {Y.}~\bibnamefont
  {Liao}}\ and\ \bibinfo {author} {\bibfnamefont {V.}~\bibnamefont
  {Galitski}},\ }\bibfield  {title} {\bibinfo {title} {Nonlinear sigma model
  approach to many-body quantum chaos: Regularized and unregularized
  out-of-time-ordered correlators},\ }\href
  {https://doi.org/10.1103/PhysRevB.98.205124} {\bibfield  {journal} {\bibinfo
  {journal} {Phys. Rev. B}\ }\textbf {\bibinfo {volume} {98}},\ \bibinfo
  {pages} {205124} (\bibinfo {year} {2018})}\BibitemShut {NoStop}%
\bibitem [{\citenamefont {Ch\'avez-Carlos}\ \emph {et~al.}(2019)\citenamefont
  {Ch\'avez-Carlos}, \citenamefont {L\'opez-del Carpio}, \citenamefont
  {Bastarrachea-Magnani}, \citenamefont {Str\'ansk\'y}, \citenamefont
  {Lerma-Hern\'andez}, \citenamefont {Santos},\ and\ \citenamefont
  {Hirsch}}]{DickeModelOTOC}%
  \BibitemOpen
  \bibfield  {author} {\bibinfo {author} {\bibfnamefont {J.}~\bibnamefont
  {Ch\'avez-Carlos}}, \bibinfo {author} {\bibfnamefont {B.}~\bibnamefont
  {L\'opez-del Carpio}}, \bibinfo {author} {\bibfnamefont {M.~A.}\ \bibnamefont
  {Bastarrachea-Magnani}}, \bibinfo {author} {\bibfnamefont {P.}~\bibnamefont
  {Str\'ansk\'y}}, \bibinfo {author} {\bibfnamefont {S.}~\bibnamefont
  {Lerma-Hern\'andez}}, \bibinfo {author} {\bibfnamefont {L.~F.}\ \bibnamefont
  {Santos}},\ and\ \bibinfo {author} {\bibfnamefont {J.~G.}\ \bibnamefont
  {Hirsch}},\ }\bibfield  {title} {\bibinfo {title} {Quantum and classical
  {L}yapunov exponents in atom-field interaction systems},\ }\href
  {https://doi.org/10.1103/PhysRevLett.122.024101} {\bibfield  {journal}
  {\bibinfo  {journal} {Phys. Rev. Lett.}\ }\textbf {\bibinfo {volume} {122}},\
  \bibinfo {pages} {024101} (\bibinfo {year} {2019})}\BibitemShut {NoStop}%
\bibitem [{\citenamefont {Pappalardi}\ \emph {et~al.}(2020)\citenamefont
  {Pappalardi}, \citenamefont {Polkovnikov},\ and\ \citenamefont
  {Silva}}]{pappalardi2020quantum}%
  \BibitemOpen
  \bibfield  {author} {\bibinfo {author} {\bibfnamefont {S.}~\bibnamefont
  {Pappalardi}}, \bibinfo {author} {\bibfnamefont {A.}~\bibnamefont
  {Polkovnikov}},\ and\ \bibinfo {author} {\bibfnamefont {A.}~\bibnamefont
  {Silva}},\ }\bibfield  {title} {\bibinfo {title} {Quantum echo dynamics in
  the {S}herrington-{K}irkpatrick model},\ }\href@noop {} {\bibfield  {journal}
  {\bibinfo  {journal} {SciPost Phys.}\ }\textbf {\bibinfo {volume} {9}},\
  \bibinfo {pages} {021} (\bibinfo {year} {2020})}\BibitemShut {NoStop}%
\bibitem [{\citenamefont {Zhang}\ \emph {et~al.}(2022)\citenamefont {Zhang},
  \citenamefont {Wolynes},\ and\ \citenamefont {Gruebele}}]{MoleculeOTOC}%
  \BibitemOpen
  \bibfield  {author} {\bibinfo {author} {\bibfnamefont {C.}~\bibnamefont
  {Zhang}}, \bibinfo {author} {\bibfnamefont {P.~G.}\ \bibnamefont {Wolynes}},\
  and\ \bibinfo {author} {\bibfnamefont {M.}~\bibnamefont {Gruebele}},\
  }\bibfield  {title} {\bibinfo {title} {Quantum information scrambling in
  molecules},\ }\href {https://doi.org/10.1103/PhysRevA.105.033322} {\bibfield
  {journal} {\bibinfo  {journal} {Phys. Rev. A}\ }\textbf {\bibinfo {volume}
  {105}},\ \bibinfo {pages} {033322} (\bibinfo {year} {2022})}\BibitemShut
  {NoStop}%
\bibitem [{\citenamefont {Sachdev}\ and\ \citenamefont
  {Ye}(1993)}]{sachdev1993gapless}%
  \BibitemOpen
  \bibfield  {author} {\bibinfo {author} {\bibfnamefont {S.}~\bibnamefont
  {Sachdev}}\ and\ \bibinfo {author} {\bibfnamefont {J.}~\bibnamefont {Ye}},\
  }\bibfield  {title} {\bibinfo {title} {Gapless spin-fluid ground state in a
  random quantum {H}eisenberg magnet},\ }\href
  {https://doi.org/10.1103/PhysRevLett.70.3339} {\bibfield  {journal} {\bibinfo
   {journal} {Phys.~Rev.~Lett.}\ }\textbf {\bibinfo {volume} {70}},\ \bibinfo
  {pages} {3339} (\bibinfo {year} {1993})}\BibitemShut {NoStop}%
\bibitem [{\citenamefont {Kitaev}()}]{KitaevSYK}%
  \BibitemOpen
  \bibfield  {author} {\bibinfo {author} {\bibfnamefont {A.}~\bibnamefont
  {Kitaev}},\ }\href {http://online.kitp.ucsb.edu/online/entangled15/kitaev/}
  {\bibinfo {title} {A simple model of quantum holography}},\ \bibinfo
  {howpublished} {talk given at KITP in 2015. See:
  \url{http://online.kitp.ucsb.edu/online/entangled15/kitaev/}}\BibitemShut
  {NoStop}%
\bibitem [{\citenamefont {Murthy}\ and\ \citenamefont
  {Srednicki}(2019)}]{murthy2019bounds}%
  \BibitemOpen
  \bibfield  {author} {\bibinfo {author} {\bibfnamefont {C.}~\bibnamefont
  {Murthy}}\ and\ \bibinfo {author} {\bibfnamefont {M.}~\bibnamefont
  {Srednicki}},\ }\bibfield  {title} {\bibinfo {title} {Bounds on chaos from
  the eigenstate thermalization hypothesis},\ }\href
  {https://doi.org/10.1103/PhysRevLett.123.230606} {\bibfield  {journal}
  {\bibinfo  {journal} {Phys. Rev. Lett.}\ }\textbf {\bibinfo {volume} {123}},\
  \bibinfo {pages} {230606} (\bibinfo {year} {2019})}\BibitemShut {NoStop}%
\bibitem [{\citenamefont {Tsuji}\ \emph {et~al.}(2018)\citenamefont {Tsuji},
  \citenamefont {Shitara},\ and\ \citenamefont {Ueda}}]{tsuji}%
  \BibitemOpen
  \bibfield  {author} {\bibinfo {author} {\bibfnamefont {N.}~\bibnamefont
  {Tsuji}}, \bibinfo {author} {\bibfnamefont {T.}~\bibnamefont {Shitara}},\
  and\ \bibinfo {author} {\bibfnamefont {M.}~\bibnamefont {Ueda}},\ }\bibfield
  {title} {\bibinfo {title} {Bound on the exponential growth rate of
  out-of-time-ordered correlators},\ }\href
  {https://doi.org/10.1103/PhysRevE.98.012216} {\bibfield  {journal} {\bibinfo
  {journal} {Phys. Rev. E}\ }\textbf {\bibinfo {volume} {98}},\ \bibinfo
  {pages} {012216} (\bibinfo {year} {2018})}\BibitemShut {NoStop}%
\bibitem [{\citenamefont {Pappalardi}\ \emph {et~al.}(2022)\citenamefont
  {Pappalardi}, \citenamefont {Foini},\ and\ \citenamefont
  {Kurchan}}]{Pappalardi2022}%
  \BibitemOpen
  \bibfield  {author} {\bibinfo {author} {\bibfnamefont {S.}~\bibnamefont
  {Pappalardi}}, \bibinfo {author} {\bibfnamefont {L.}~\bibnamefont {Foini}},\
  and\ \bibinfo {author} {\bibfnamefont {J.}~\bibnamefont {Kurchan}},\
  }\bibfield  {title} {\bibinfo {title} {Quantum bounds and
  fluctuation-dissipation relations},\ }\href@noop {} {\bibfield  {journal}
  {\bibinfo  {journal} {SciPost Phys.}\ }\textbf {\bibinfo {volume} {12}},\
  \bibinfo {pages} {130} (\bibinfo {year} {2022})}\BibitemShut {NoStop}%
\bibitem [{\citenamefont {Nussinov}\ and\ \citenamefont
  {Chakrabarty}(2022)}]{nussinov2022exact}%
  \BibitemOpen
  \bibfield  {author} {\bibinfo {author} {\bibfnamefont {Z.}~\bibnamefont
  {Nussinov}}\ and\ \bibinfo {author} {\bibfnamefont {S.}~\bibnamefont
  {Chakrabarty}},\ }\bibfield  {title} {\bibinfo {title} {Exact universal
  chaos, speed limit, acceleration, planckian transport
  coefficient,“collapse” to equilibrium, and other bounds in thermal
  quantum systems},\ }\href@noop {} {\bibfield  {journal} {\bibinfo  {journal}
  {Annals of Physics}\ }\textbf {\bibinfo {volume} {443}},\ \bibinfo {pages}
  {168970} (\bibinfo {year} {2022})}\BibitemShut {NoStop}%
\bibitem [{\citenamefont {Parrinello}\ and\ \citenamefont
  {Rahman}(1984)}]{rahman}%
  \BibitemOpen
  \bibfield  {author} {\bibinfo {author} {\bibfnamefont {M.}~\bibnamefont
  {Parrinello}}\ and\ \bibinfo {author} {\bibfnamefont {A.}~\bibnamefont
  {Rahman}},\ }\bibfield  {title} {\bibinfo {title} {Study of an {F} center in
  molten {KC}l},\ }\href@noop {} {\bibfield  {journal} {\bibinfo  {journal} {J.
  Chem. Phys.}\ }\textbf {\bibinfo {volume} {80}},\ \bibinfo {pages} {860}
  (\bibinfo {year} {1984})}\BibitemShut {NoStop}%
\bibitem [{\citenamefont {Cao}\ and\ \citenamefont {Voth}(1994)}]{voth}%
  \BibitemOpen
  \bibfield  {author} {\bibinfo {author} {\bibfnamefont {J.}~\bibnamefont
  {Cao}}\ and\ \bibinfo {author} {\bibfnamefont {G.~A.}\ \bibnamefont {Voth}},\
  }\bibfield  {title} {\bibinfo {title} {The formulation of quantum statistical
  mechanics based on the {F}eynman path centroid density. {IV}. {A}lgorithms
  for centroid molecular dynamics},\ }\href@noop {} {\bibfield  {journal}
  {\bibinfo  {journal} {J. Chem. Phys.}\ }\textbf {\bibinfo {volume} {101}},\
  \bibinfo {pages} {6168} (\bibinfo {year} {1994})}\BibitemShut {NoStop}%
\bibitem [{\citenamefont {Craig}\ and\ \citenamefont
  {Manolopoulos}(2004)}]{craig2004quantum}%
  \BibitemOpen
  \bibfield  {author} {\bibinfo {author} {\bibfnamefont {I.~R.}\ \bibnamefont
  {Craig}}\ and\ \bibinfo {author} {\bibfnamefont {D.~E.}\ \bibnamefont
  {Manolopoulos}},\ }\bibfield  {title} {\bibinfo {title} {Quantum statistics
  and classical mechanics: Real time correlation functions from ring polymer
  molecular dynamics},\ }\href@noop {} {\bibfield  {journal} {\bibinfo
  {journal} {J. Chem. Phys.}\ }\textbf {\bibinfo {volume} {121}},\ \bibinfo
  {pages} {3368} (\bibinfo {year} {2004})}\BibitemShut {NoStop}%
\bibitem [{\citenamefont {Habershon}\ \emph {et~al.}(2013)\citenamefont
  {Habershon}, \citenamefont {Manolopoulos}, \citenamefont {Markland},\ and\
  \citenamefont {Miller}}]{annurev}%
  \BibitemOpen
  \bibfield  {author} {\bibinfo {author} {\bibfnamefont {S.}~\bibnamefont
  {Habershon}}, \bibinfo {author} {\bibfnamefont {D.~E.}\ \bibnamefont
  {Manolopoulos}}, \bibinfo {author} {\bibfnamefont {T.~E.}\ \bibnamefont
  {Markland}},\ and\ \bibinfo {author} {\bibfnamefont {T.~F.}\ \bibnamefont
  {Miller}},\ }\bibfield  {title} {\bibinfo {title} {Ring-polymer molecular
  dynamics: Quantum effects in chemical dynamics from classical trajectories in
  an extended phase space},\ }\href@noop {} {\bibfield  {journal} {\bibinfo
  {journal} {Annu. Rev. Phys. Chem.}\ }\textbf {\bibinfo {volume} {64}},\
  \bibinfo {pages} {387} (\bibinfo {year} {2013})}\BibitemShut {NoStop}%
\bibitem [{\citenamefont {Hashimoto}\ \emph {et~al.}(2020)\citenamefont
  {Hashimoto}, \citenamefont {Huh}, \citenamefont {Kim},\ and\ \citenamefont
  {Watanabe}}]{hashimoto2020exponential}%
  \BibitemOpen
  \bibfield  {author} {\bibinfo {author} {\bibfnamefont {K.}~\bibnamefont
  {Hashimoto}}, \bibinfo {author} {\bibfnamefont {K.-B.}\ \bibnamefont {Huh}},
  \bibinfo {author} {\bibfnamefont {K.-Y.}\ \bibnamefont {Kim}},\ and\ \bibinfo
  {author} {\bibfnamefont {R.}~\bibnamefont {Watanabe}},\ }\bibfield  {title}
  {\bibinfo {title} {Exponential growth of out-of-time-order correlator without
  chaos: inverted harmonic oscillator},\ }\href@noop {} {\bibfield  {journal}
  {\bibinfo  {journal} {J. High Energy Phys.}\ }\textbf {\bibinfo {volume}
  {2020}}\bibinfo  {number} { (11)},\ \bibinfo {pages} {1}}\BibitemShut
  {NoStop}%
\bibitem [{\citenamefont {Braams}\ and\ \citenamefont
  {Manolopoulos}(2006)}]{braams}%
  \BibitemOpen
\bibfield  {number} {  }\bibfield  {author} {\bibinfo {author} {\bibfnamefont
  {B.~J.}\ \bibnamefont {Braams}}\ and\ \bibinfo {author} {\bibfnamefont
  {D.~E.}\ \bibnamefont {Manolopoulos}},\ }\bibfield  {title} {\bibinfo {title}
  {Ring-polymer molecular dynamics: Quantum effects in chemical dynamics from
  classical trajectories in an extended phase space},\ }\href@noop {}
  {\bibfield  {journal} {\bibinfo  {journal} {J. Phys. Chem.}\ }\textbf
  {\bibinfo {volume} {125}},\ \bibinfo {pages} {124105} (\bibinfo {year}
  {2006})}\BibitemShut {NoStop}%
\bibitem [{\citenamefont {Brewer}\ \emph {et~al.}(1997)\citenamefont {Brewer},
  \citenamefont {Hulme},\ and\ \citenamefont
  {Manolopoulos}}]{brewer1997semiclassical}%
  \BibitemOpen
  \bibfield  {author} {\bibinfo {author} {\bibfnamefont {M.~L.}\ \bibnamefont
  {Brewer}}, \bibinfo {author} {\bibfnamefont {J.~S.}\ \bibnamefont {Hulme}},\
  and\ \bibinfo {author} {\bibfnamefont {D.~E.}\ \bibnamefont {Manolopoulos}},\
  }\bibfield  {title} {\bibinfo {title} {Semiclassical dynamics in up to 15
  coupled vibrational degrees of freedom},\ }\href@noop {} {\bibfield
  {journal} {\bibinfo  {journal} {J. Chem. Phys.}\ }\textbf {\bibinfo {volume}
  {106}},\ \bibinfo {pages} {4832} (\bibinfo {year} {1997})}\BibitemShut
  {NoStop}%
\bibitem [{Note1()}]{Note1}%
  \BibitemOpen
  \bibinfo {note} {The oscillations in the post-Ehrenfest part of the quantum
  OTOC at $0.95T_c$ demonstrate that the system has not completely thermalised
  after scrambling. Similar differences in the scrambling and thermalization
  timescales have been found in calculations on the Dicke model \cite
  {DickeModelOTOC}.}\BibitemShut {Stop}%
\bibitem [{Note2()}]{Note2}%
  \BibitemOpen
  \bibinfo {note} {Note that the quantum Kubo OTOC of Eq.~(\ref {kubo}) is
  different from the various types of symmetrised quantum OTOC usually
  considered when discussing the bound \cite {RegOTOC}. However, like the
  symmetrised OTOCs, the quantum Kubo OTOC is `regularised', with the two
  commutators separated by ${\exp }(-\gamma \protect \hat H)$.}\BibitemShut
  {Stop}%
\bibitem [{\citenamefont {Miller}(1975)}]{miller}%
  \BibitemOpen
  \bibfield  {author} {\bibinfo {author} {\bibfnamefont {W.~H.}\ \bibnamefont
  {Miller}},\ }\bibfield  {title} {\bibinfo {title} {Semiclassical limit of
  quantum mechanical transition state theory for nonseparable systems},\
  }\href@noop {} {\bibfield  {journal} {\bibinfo  {journal} {J. Chem. Phys.}\
  }\textbf {\bibinfo {volume} {62}},\ \bibinfo {pages} {1899} (\bibinfo {year}
  {1975})}\BibitemShut {NoStop}%
\bibitem [{\citenamefont {Benderskii}\ \emph {et~al.}(1994)\citenamefont
  {Benderskii}, \citenamefont {E},\ and\ \citenamefont {A}}]{benders}%
  \BibitemOpen
  \bibfield  {author} {\bibinfo {author} {\bibfnamefont {V.~A.}\ \bibnamefont
  {Benderskii}}, \bibinfo {author} {\bibfnamefont {M.~D.}\ \bibnamefont {E}},\
  and\ \bibinfo {author} {\bibfnamefont {W.~C.}\ \bibnamefont {A}},\ }\bibfield
   {title} {\bibinfo {title} {Chemical dynamics at low temperatures},\
  }\href@noop {} {\bibfield  {journal} {\bibinfo  {journal} {Adv. Chem. Phys.}\
  }\textbf {\bibinfo {volume} {88}},\ \bibinfo {pages} {55} (\bibinfo {year}
  {1994})}\BibitemShut {NoStop}%
\bibitem [{\citenamefont {Richardson}\ and\ \citenamefont
  {Althorpe}(2009)}]{richardson2009ring}%
  \BibitemOpen
  \bibfield  {author} {\bibinfo {author} {\bibfnamefont {J.~O.}\ \bibnamefont
  {Richardson}}\ and\ \bibinfo {author} {\bibfnamefont {S.~C.}\ \bibnamefont
  {Althorpe}},\ }\bibfield  {title} {\bibinfo {title} {Ring-polymer molecular
  dynamics rate-theory in the deep-tunneling regime: Connection with
  semiclassical instanton theory},\ }\href@noop {} {\bibfield  {journal}
  {\bibinfo  {journal} {J. Chem. Phys.}\ }\textbf {\bibinfo {volume} {131}},\
  \bibinfo {pages} {214106} (\bibinfo {year} {2009})}\BibitemShut {NoStop}%
\bibitem [{\citenamefont {Hashimoto}\ and\ \citenamefont
  {Tanahashi}(2017)}]{HawkingTempHashimoto}%
  \BibitemOpen
  \bibfield  {author} {\bibinfo {author} {\bibfnamefont {K.}~\bibnamefont
  {Hashimoto}}\ and\ \bibinfo {author} {\bibfnamefont {N.}~\bibnamefont
  {Tanahashi}},\ }\bibfield  {title} {\bibinfo {title} {Universality in chaos
  of particle motion near black hole horizon},\ }\href
  {https://doi.org/10.1103/PhysRevD.95.024007} {\bibfield  {journal} {\bibinfo
  {journal} {Phys. Rev. D}\ }\textbf {\bibinfo {volume} {95}},\ \bibinfo
  {pages} {024007} (\bibinfo {year} {2017})}\BibitemShut {NoStop}%
\bibitem [{Note3()}]{Note3}%
  \BibitemOpen
  \bibinfo {note} {These OTOCs are of course not microcanonical except in the
  high-temperature limit (since $H_N$ is not the energy); they are used here
  simply to probe the dynamics of the trajectories with energy close to the
  barrier top.}\BibitemShut {Stop}%
\bibitem [{\citenamefont {Hele}\ and\ \citenamefont {Althorpe}(2013)}]{tim}%
  \BibitemOpen
  \bibfield  {author} {\bibinfo {author} {\bibfnamefont {T.~J.~H.}\
  \bibnamefont {Hele}}\ and\ \bibinfo {author} {\bibfnamefont {S.~C.}\
  \bibnamefont {Althorpe}},\ }\bibfield  {title} {\bibinfo {title} {Derivation
  of a true quantum transition-state theory. {I}. {U}niqueness and equivalence
  to ring-polymer molecular dynamics transition-state-theory},\ }\href@noop {}
  {\bibfield  {journal} {\bibinfo  {journal} {J. Chem. Phys.}\ }\textbf
  {\bibinfo {volume} {138}},\ \bibinfo {pages} {084108} (\bibinfo {year}
  {2013})}\BibitemShut {NoStop}%
\bibitem [{\citenamefont {Kidd}\ \emph {et~al.}(2021)\citenamefont {Kidd},
  \citenamefont {Safavi-Naini},\ and\ \citenamefont {Corney}}]{SPscramNotherm}%
  \BibitemOpen
  \bibfield  {author} {\bibinfo {author} {\bibfnamefont {R.~A.}\ \bibnamefont
  {Kidd}}, \bibinfo {author} {\bibfnamefont {A.}~\bibnamefont {Safavi-Naini}},\
  and\ \bibinfo {author} {\bibfnamefont {J.~F.}\ \bibnamefont {Corney}},\
  }\bibfield  {title} {\bibinfo {title} {Saddle-point scrambling without
  thermalization},\ }\href {https://doi.org/10.1103/PhysRevA.103.033304}
  {\bibfield  {journal} {\bibinfo  {journal} {Phys. Rev. A}\ }\textbf {\bibinfo
  {volume} {103}},\ \bibinfo {pages} {033304} (\bibinfo {year}
  {2021})}\BibitemShut {NoStop}%
\bibitem [{\citenamefont {Kidd}\ \emph {et~al.}(2020)\citenamefont {Kidd},
  \citenamefont {Safavi-Naini},\ and\ \citenamefont
  {Corney}}]{kidd2020thermalization}%
  \BibitemOpen
  \bibfield  {author} {\bibinfo {author} {\bibfnamefont {R.~A.}\ \bibnamefont
  {Kidd}}, \bibinfo {author} {\bibfnamefont {A.}~\bibnamefont {Safavi-Naini}},\
  and\ \bibinfo {author} {\bibfnamefont {J.~F.}\ \bibnamefont {Corney}},\
  }\bibfield  {title} {\bibinfo {title} {Thermalization in a bose-hubbard dimer
  with modulated tunneling},\ }\href
  {https://doi.org/10.1103/PhysRevA.102.023330} {\bibfield  {journal} {\bibinfo
   {journal} {Phys. Rev. A}\ }\textbf {\bibinfo {volume} {102}},\ \bibinfo
  {pages} {023330} (\bibinfo {year} {2020})}\BibitemShut {NoStop}%
\bibitem [{\citenamefont {Fu}\ and\ \citenamefont
  {Sachdev}(2016)}]{fu2016numerical}%
  \BibitemOpen
  \bibfield  {author} {\bibinfo {author} {\bibfnamefont {W.}~\bibnamefont
  {Fu}}\ and\ \bibinfo {author} {\bibfnamefont {S.}~\bibnamefont {Sachdev}},\
  }\bibfield  {title} {\bibinfo {title} {Numerical study of fermion and boson
  models with infinite-range random interactions},\ }\href
  {https://doi.org/10.1103/PhysRevB.94.035135} {\bibfield  {journal} {\bibinfo
  {journal} {Phys. Rev. B}\ }\textbf {\bibinfo {volume} {94}},\ \bibinfo
  {pages} {035135} (\bibinfo {year} {2016})}\BibitemShut {NoStop}%
\bibitem [{\citenamefont {Colbert}\ and\ \citenamefont
  {Miller}(1992)}]{colbert1992novel}%
  \BibitemOpen
  \bibfield  {author} {\bibinfo {author} {\bibfnamefont {D.~T.}\ \bibnamefont
  {Colbert}}\ and\ \bibinfo {author} {\bibfnamefont {W.~H.}\ \bibnamefont
  {Miller}},\ }\bibfield  {title} {\bibinfo {title} {A novel discrete variable
  representation for quantum mechanical reactive scattering via the s-matrix
  kohn method},\ }\href@noop {} {\bibfield  {journal} {\bibinfo  {journal} {The
  Journal of chemical physics}\ }\textbf {\bibinfo {volume} {96}},\ \bibinfo
  {pages} {1982} (\bibinfo {year} {1992})}\BibitemShut {NoStop}%
\bibitem [{\citenamefont {Ceriotti}\ \emph {et~al.}(2010)\citenamefont
  {Ceriotti}, \citenamefont {Parrinello}, \citenamefont {Markland},\ and\
  \citenamefont {Manolopoulos}}]{ceriotti2010efficient}%
  \BibitemOpen
  \bibfield  {author} {\bibinfo {author} {\bibfnamefont {M.}~\bibnamefont
  {Ceriotti}}, \bibinfo {author} {\bibfnamefont {M.}~\bibnamefont
  {Parrinello}}, \bibinfo {author} {\bibfnamefont {T.~E.}\ \bibnamefont
  {Markland}},\ and\ \bibinfo {author} {\bibfnamefont {D.~E.}\ \bibnamefont
  {Manolopoulos}},\ }\bibfield  {title} {\bibinfo {title} {Efficient stochastic
  thermostatting of path integral molecular dynamics},\ }\href@noop {}
  {\bibfield  {journal} {\bibinfo  {journal} {The Journal of chemical physics}\
  }\textbf {\bibinfo {volume} {133}},\ \bibinfo {pages} {124104} (\bibinfo
  {year} {2010})}\BibitemShut {NoStop}%
\end{thebibliography}%

\onecolumngrid
\pagebreak
\widetext
\begin{center}
\textbf{\large Instantons and the quantum bound to chaos:\\ Supplementary information}
\end{center}
\setcounter{equation}{0}
\setcounter{figure}{0}
\setcounter{table}{0}
\setcounter{page}{1}
\makeatletter
\renewcommand{\theequation}{S\arabic{equation}}
\renewcommand{\thefigure}{S\arabic{figure}}
\renewcommand{\bibnumfmt}[1]{[S#1]}
\renewcommand{\citenumfont}[1]{S#1}

\large

\section{Methods}
\subsection{Exact quantum calculations}
The quantum Kubo OTOC of eqn. 10 in the main text was calculated using the expansion
\begin{align*}
    C(t) 
    &=-\frac{1}{\beta Z} \sum_{n,m=1}^{\infty}  \frac{e^{-\beta E_n} - e^{-\beta E_m}}{E_n - E_m} \abs{\bra{n} [\hat{W}(t),\hat{V}] \ket{m}}^2
\end{align*}
in which the hamiltonian eigenstates $\ket{n}$ were calculated using the grid basis of ref.~\cite{colbert1992novel}.

The Husimi distribution in Fig.~1(d) is given by
\begin{align}
    \mathcal{H}^{\pm}_n(x,p_x; y_0) =  \abs{\bra{\textbf{p}_0,\textbf{q}_0}\ket{n_x = 3, n_y = 2}}^2
 \end{align}
where
\begin{align}
    \bra{\textbf{q}}\ket{\textbf{p}_0,\textbf{q}_0} = \left (\frac{1}{\pi \hbar \Sigma} \right)^{\sfrac{1}{4}} \exp{-\frac{1}{2\hbar\Sigma}  \lvert\textbf{q}- \textbf{q}_0\rvert^2  + \frac{i}{\hbar}\:{\textbf{p}_0 \cdot (\textbf{q}- \textbf{q}_0) }} 
\end{align}
is a coherent state of width $\Sigma=1.0$, ${\textbf{q}_0} = (x,0)$ and
 ${\textbf{p}_0} = (p_x,p_y)$, with $p_y= \sqrt{2m [E_{n_x = 3, n_y = 2} - V(x,0)] - p_x^2}$.

\subsection{RPMD calculations}

The RPMD OTOCs of Fig.~2 were computed using $N=16$ polymer beads at $T=3T_c,0.95T_c$, and $N= 32$ at $T=0.6T_c$. The ring-polymer distributions were thermalised by attaching a path-integral Langevin equation (PILE) thermostat  \cite{ceriotti2010efficient} and propagating for 50 atomic units of time, using the velocity-Verlet algorithm with a timestep of 0.05 a.u. A sample of 100,000 ring-polymer geometries were chosen at random from the equilibrium distributions, the stability (monodromy) matrices of which were then propagated for 5 atomic units of time using the fourth order symplectic integrator of ref. \cite{brewer1997semiclassical} with a timestep of 0.002 a.u.

\subsection{Classical and centroid Poincar\'e surfaces of section}

The classical Poincar\'e surfaces of section (PSOS) were obtained by initializing classical trajectories at random points in phase space with an energy $E$, then allowing the trajectories to explore phase space for 50 a.u.\ of time, after which the PSOSs were generated by following the trajectories for a further 200--1000~a.u. All classical PSOSs were computed with $y_0 = 0$ (and $p_y>0$) in order to visualize the dynamics around the saddle point.

\vspace{5pt}

The centroid surfaces of section (CSOS) were obtained by initializing RPMD trajectories at random points in RPMD phase-space with a total energy $NE$ (where $N$ is the number of beads), then allowing the trajectories to explore phase space for 100 a.u.\ of time. The CSOSs were then generated by plotting $X_0$ and $P_{X,0}$ (the centroids of the $x$-coordinates of the beads and their conjugate momenta) whenever the ring-polymer crosses the $Y_0 = 0$ hyperplane  with $P_{Y,0}>0$; this hyperplane is of dimension $4N-2$, so the CSOS is just a convenient 2-dimensional projection of the hyperplane, which tends to the classical PSOS in the high-temperature limit (where the polymer beads collapse to a point). The same 100 a.u.-long trajectories were also used to initialize the propagation of the stability matrix that were used to generate the `microcanonical' RPMD OTOCs of eqn 16 in the main text. 

\subsection{Histograms of the ring-polymer radius of gyration}
The radius of gyration $r_g$ of a ring polymer with respect to the centroid is defined to be
\begin{align}
    r_g = \abs{\frac{\sum_{i=1}^N \norm{\textbf{q}_i - \textbf{Q}_0}^2}{N}}^{\sfrac{1}{2}}
\end{align}
The value of $r_g$ changes along a ring-polymer trajectory, typically increasing near the barrier top. The histogram in Fig.~3c plots the distribution of the maximum values of $r_g$ along each set of ring-polymer trajectories propagated for 1000 a.u.\ of time.

\section*{Animations of RPMD trajectories}
The full RPMD trajectory from which the snapshots of Fig.~3e were taken can be viewed at \href{https://drive.google.com/file/d/1h2aWPPTjr0oXgqU7VNi_u_QnrliHvJCH/view?usp=share_link}{\textbf{Long animation} } and a short section cut from it at \href{https://drive.google.com/file/d/1jLMPzRK4DmUkoNP_SkeewNA5kKDZbfVk/view?usp=share_link}{\textbf{Short animation}}

\end{document}